\documentclass[runningheads]{llncs}
\usepackage{graphicx}
\usepackage{amsmath}

\begin{document}
\title{Fully Convolutional Network Bootstrapped by Word Encoding and Embedding for Activity Recognition in Smart Homes\thanks{The work is partially supported by project VITAAL and is financed by Brest Metropole, the region of Brittany and the European Regional Development Fund (ERDF).}}
\author{Damien Bouchabou\inst{1,2} \and
Sao Mai Nguyen\inst{1} \orcidID{0000-0003-0929-0019}\and
Christophe Lohr\inst{1} \and
Benoit LeDuc\inst{2}\and
Ioannis Kanellos\inst{1}}
\authorrunning{D. Bouchabou et al.}
\institute{IMT Atlantique, Lab-STICC, UMR 6285, F-29238 Brest, France 
\email{\{damien.bouchabou, christophe.lohr, ioannis.kanellos\}@imt-atlantique.fr}\\
\and
Delta Dore company, Bonnemain, France
\email{\{dbouchabou, bleduc\}@deltadore.com}\\
% \and
% FLOWERS team, U2IS, ENSTA, IP Paris \& Inria, France\\
\email{\{nguyensmai\}@gmail.com}}
\maketitle              % typeset the header of the contribution
\begin{abstract}
Activity recognition in smart homes is essential when we wish to propose automatic services for the inhabitants. However, it is a challenging problem in terms of environments’ variability, sensory-motor systems, user habits, but also sparsity of signals and redundancy of models . Therefore, end-to-end systems fail at automatically extracting key features, and need to access context and domain knowledge. We propose to tackle feature extraction for activity recognition in smart homes by merging methods of Natural Language Processing (NLP) and Time Series Classification (TSC) domains. 

We evaluate the performance of our method with two datasets issued from the Center for Advanced Studies in Adaptive Systems (CASAS).
We analyze the contributions of the use of embedding based on term frequency encoding, to improve automatic feature extraction. Moreover we compare the classification performance of Fully Convolutional Network (FCN) from TSC, applied for the first time for activity recognition in smart homes, to Long Short Term Memory (LSTM). The method we propose, shows good performance in offline activity classification. Our analysis also shows that FCNs outperforms LSTMs, and that domain knowledge gained by event encoding and embedding improves significantly the performance of classifiers.

\end{abstract}
\section{Introduction}
Human Activity Recognition (HAR) has been the focus of research efforts due to its key role for different ambient assisted living (AAL) domains as well as the increasing demand for home automation and convenience services in daily activities. The main task of HAR is to recognize human activities from the data collected through environmental sensors and Internet of Things (IoT) devices. They use different sensor technologies such as cameras, wearable or low-level smart sensors to track human activities, as described in \cite{hussain2019different}.

Recent advances in IoT technologies and the reduction of the cost of sensors are leading to the proliferation of these ambient devices and the development of smart homes. This is why in this work we will focus more on IoT-based HAR, as opposed to video or wearable-based HAR.

Along the development of the hardware, the HAR algorithms also need to solve the challenges of HAR in smart homes. % is not an easy problem. % indeed human activity is much more complex than such disjoint occurrences of simple actions \cite{liu2016action2activity}. The actions underlying an activity are not independent. Sophisticated temporal combinations lead to semantic differences in activity understanding. Commonality or semantic relatedness exist across multiple activities. Moreover, 
Indeed,  the number, the type but also the placement of sensors can significantly influence the performance of HAR systems. A system suitable for a given home may be completely inadequate in some other, due to different house configuration or user habits. The algorithms thus need to be robust to the variability of environments. 
Besides, while video-based HAR can leverage rich and redundant information from images and video streams, IoT based HAR faces the challenges of sparse and incomplete information and redundant models. In contrast to videos where objects and people appear on several pixels and over several video frames, the IoT network only detects changes in the environment that are within their range of detection and in their field of view, and is oblivious to most changes in the environment, which occur outside these ranges. When a change is captured, this detection often translates into a signal with a single value from one sensor. This sparsity entails the redundancy challenge: a set of signals from the same set of sensors can be caused by different activities.  % Besides, the activities are not simply identifiable from the id of the activated sensor.%different activities can be captured by the same sensors.
Thus, algorithms for HAR in smart homes need to address the challenges of variability, sparsity and redundancy.
 
To adapt to variations of environments and uses, algorithms for HAR have turned to machine learning methods, and more specifically Deep Learning (DL) algorithms. To deal with sparsity and redundancy, first, algorithms that can learn long-term dependencies have been developed so as to understand the context of sensor signals. Second, studies have also tried to introduce domain knowledge and contextualization of sensors signals, through a good feature representation of sensor events. But handcrafted features need a lot of pre-processing and reduce its adaptability to various environments. Therefore HAR algorithms need to automatically extract domain relevant representations.

In recent years, there have been significant improvements of DL techniques. They have been successfully applied to Natural Language Processing (NLP) and Time Series Classification (TSC). Respectively for automatic extraction of good feature representations through word embedding techniques and classifiers.

Our contributions are the following: 1) We apply for the first time the Fully Convolutional  Networks (FCN) classifier from TSC on activity recognition in smart homes. 2) We propose to use frequency-based encoding with word embedding from NLP to improve automatic feature extraction. 3) We design an end to end framework to automatically extract key features and classify daily activities in smart homes by merging TSC classifier and NLP words encoding.
4) Finally, we show that domain knowledge gained by event encoding and embedding improves significantly the performance of classifiers.

We propose in the following section to review the state-of-the-art HAR smart home classifiers, a TSC classifier and the existing feature representation methods, in particular those used in NLP applications. In Section \ref{sec:methodology}, we will propose a framework combining a TSC algorithm and a NLP sequence features extractor method. In Section \ref{sec:results}, we will report on the performance of our proposed framework before concluding.

\section{Related Works} \label{related_wk}
%As mentioned in the introduction, we use the frequency Bag-Of-Word and the Embedding technique coupled with FCNs. But first, in this section we will describe the traditional HAR approaches. Then we will explain DL techniques and Finally we will describe the coupling between NLP encoding and a TSC classification algorithm. 
In this section, we describe the algorithms developed for HAR, and more generally for Time Serie Classification. We then examine how TSC can be bootstrapped by incorporating domain knowledge in feature encoding as in Natural Language Processing.

\subsection{Traditional HAR Approaches}

To recognize human activities based on sensor traces, researchers used various machine learning algorithms as reviewed in \cite{sedky2018evaluating}. These can be divided into two streams: the algorithms exploiting a spatiotemporal representation, with Naive Bayes, Dynamic Bayesian Networks, Hidden Markov Models; % for their ability to represent spatio-temporal information. 
and the algorithms based on features classification, with Decision Tree, Support Vector Machines, or Conditional Random Fields. % to try to find the boundary between different activity classes based on different thresholds or properties of the associated features.

Most of these traditional HAR approaches commonly use handcrafted feature extraction methods. Automatic feature extraction is one of the challenges addressed by DL.

\subsection{Deep Learning Approaches} \label{dl_approches}
Recently, a variety of DL algorithms have been applied for HAR to overcome those limitations and improve the performance of HAR. DL methods learn the features directly from the raw data hierarchically, % which eliminates the problem of handcrafted feature approximations. DL largely relieves the effort on designing features and can learn much more 
to uncover high-level features. Long Short Term Memory (LSTM) can be seen as a very successful extension of the Recurrent Neural Networks (RNN), explicitly designed to deal with long-term dependencies. LSTMs allow automatic learning of temporal information from the sensor data without the need of handcrafted features or kernel fusion approaches, and have led to good performance in HAR in smart homes, as reported in \cite{singh2017human,liciotti_lstm}. 
\cite{liciotti_lstm} evaluated different LSTMs structures for HAR in smart homes. They show that the LSTM approach outperforms traditional HAR approaches in terms of classification score without using handcrafted features. LSTMs leads to a viable solution to significantly improve the HAR task in the smart home but suffers from training time.

Another DL approach, focusing more on pattern detection is Convolutional Neural Networks (CNNs). They have three advantages for HAR. They can capture local dependencies, that is, the importance of neighboring observations correlated with the current event. They are scale invariant in terms of step difference or event frequencies. In addition, they are able to learn a hierarchical representation of data. 
Researchers used 2D \cite{gochoo2018unobtrusive,mohmed2020employing} and 1D \cite{singh2017convolutional} CNNs on HAR in smart homes. The 2D CNN obtained good classification results. But this approach is not robust enough to deal with unbalanced datasets, unlabeled events, and is not suitable for online recognition. 1D CNNs are competitive with LSTMs on sequence problems \cite{singh2017convolutional}. In general LSTMs obtain better performances due to their capacity to use long-term dependencies. But CNNs are faster to train and get accuracy levels close to LSTMs.

The FCN is a particular CNN, with only convolutional layers and no dense layers. FCN has shown compelling quality and efficiency for semantic segmentation of images \cite{long2015fully}. %Recently, some effort has been spent to exploit deep neural networks, especially CNNs for end-to-end TSC.
Due to its performance on feature extraction, researchers transferred the FCN on TSC problems \cite{wang2017time}. \cite{fawaz2019deep} compared the FCN against other TSC algorithms and obtained high classification performances. FCNs ranked first on 18 datasets out of 97 and in the top five on the others. However, no application of FCN for HAR in smart homes has been reported. For this reason we propose to apply the FCN to HAR in smart homes as a high-level extractor of features and classifier.

\subsection{NLP and TSC coupling}
Works such as \cite{tahir2020key,yan2019using} have shown the importance of a good feature representation, but designing features for HAR applications is a tedious task. %While DL algorithms can automatically extract features, % compares different features generation techniques and HAR algorithms. Their experiments compare
%the comparison in \cite{yan2019using} for HAR of handcrafted features, handcrafted with probabilistic features and DL extracted features as inputs to LSTMs and CNNs, %, in sliding windows for HAR on the sensor's streaming data. The 
% shows that the best performance is achieved by using handcrafted with probabilistic features and the Random Forest classifier. These results 
%the comparative study in \cite{yan2019using} for HAR indicates that the features generated by CNN and LSTM can be outperformed by handcrafted features.  
%the performance of DL algorithms can be improved by pre-processing the input data. Indeed as their results show  better performance by LSTMs and CNNs when using handcrafted with probabilistic features than simple handcrafted features or DL extracted features. %Authors don't use DL algorithms on raw data but on basic handcrafted features.

DL algorithms can automatically extract features, they have widely shown to improve feature representation with words pre-processing for text classification in NLP. % is an essential and significant task. The main task is to assign a label to a text or a sentence.
%This challenging problem needs efficient features extraction techniques and accurate classifiers, as in HAR.
Researchers have devised many language models and different encoding of words. % as inputs to classifiers. 
They proposed encodings such as n-gram, term frequency, term frequency-inverse document frequency, bag-of-words. Recently, they use DL algorithms such as word2vec, GloVe ELMo and more recently Transformers, coupled with the aforementioned encoding to achieve meaning word encoding \cite{kowsari2019text,li2020survey}. DL algorithms infer features from the current input and to a lesser extent from past inputs, these encodings incorporate more general domain knowledge from the whole corpus. % All these encoding methods allow great advances in NLP.
Their strong capacity to generate features from raw data and model word sequences increases the performance of DL classifiers. %HAR in smart homes can be seen as an event sequence classification problem similar to NLP text classification.
We propose to transpose previously cited NLP techniques on smart homes HAR problems in order to automatically generate key features.

Thus, we introduce in this article a DL methodology for HAR in smart homes inspired by the NLP and the TSC. We propose to combine the term frequency encoding and embedding with FCN, respectively : incorporate domain knowledge of event encoding in the first level of extraction features ; and realize a higher-level of extraction features and the activity classification. The choice of the FCN algorithm from TSC is led by the output of the NLP embedding, which transforms the sequence classification problem into a multivariate TSC problem. To our knowledge this is the first time that a study has used FCN in smart-home activity recognition, and has combined it with embedding techniques to perform an end-to-end system that automatically extracts key features and classifies activities in smart homes.

\section{Methodology} \label{sec:methodology}
We merge NLP encoding and FCN classifier from TSC to deal with smart homes HAR. This coupling allows generating automatic key features and classify activities. 

\begin{figure}[htb]
    \centering
    \includegraphics[scale=0.85]{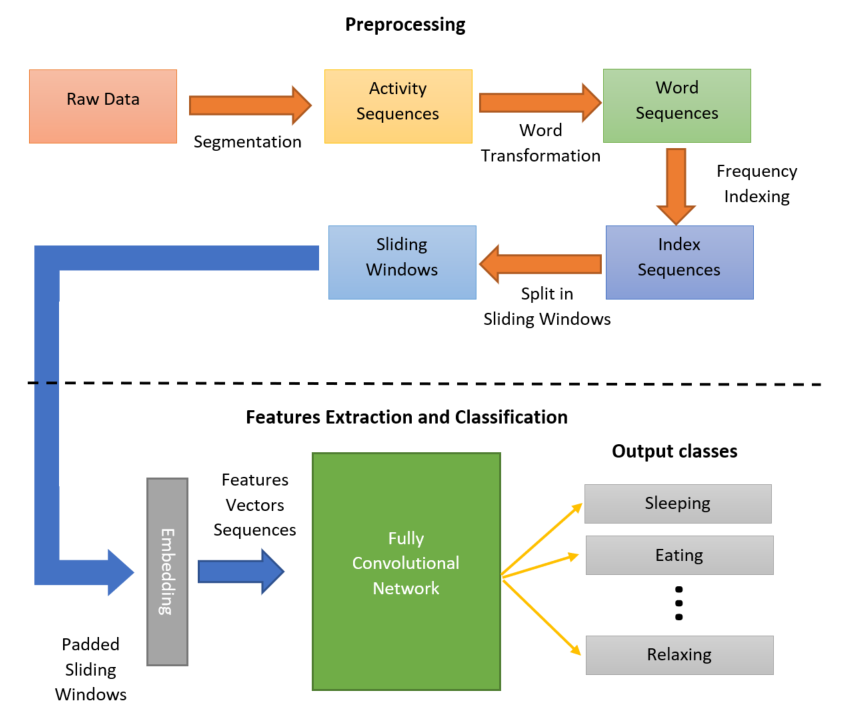}
    \caption{Framework architecture of the proposed method}
    \label{fig:FCN_framework}
\end{figure}

The framework architecture of the proposed method is shown in Figure~\ref{fig:FCN_framework}. First raw data from sensors are encoded into a sequence of indexes (section \ref{nlp_pre_process}), then are split using a sliding window (section \ref{sliding_windows}). 
The sliding windows are then processed through an embedding to extract a first level of features, and  finally classified by the FCN (section \ref{sec:fcn}).

\subsection{Problem Definition}
The activity recognition problem is a classification problem. The goal is to attribute an activity label on sensors events sequences. We model our problem as follows. A set of sensors $S = \{s_1,...,s_{|S|}\}$ produces events $e_i \in E $. An event is the value or the state returned by a sensor when the sensors emit a signal: $e_i = (s_i,v_i,t_i)$, where $s_i$ is the sensor id, $v_i$ the value returned by the sensor and $t_i$ the time when the sensor changes its state or value. A sequence $L_i$ is a trace of activity. $L_i$ is a list of events $Seq_i = (e_i, ..., e_n)$. Each $L_i$ can be associated to an activity label $a_i \in A$, by semantic segmentation.

In this paper we did not take in consideration the timestamp $t_i$ when an event occurs. We simply ignore the parameter $t$ for our experiments. We want to be able to recognize an activity regardless of the time of the day. For example, the activity "Sleeping" appears in general during the night but this activity can appear at any time during the day. Some people can work during the day and sleep by night and vice versa some people can work during the night and sleep during the day. 

\subsection{NLP Encoding} \label{nlp_pre_process}
Our hypothesis is to process sensor events like words and activity sequences as text sentences; these sentences describing the activities carried out by the inhabitants.

First, each sequence of activity is extracted from the dataset as sentences in NLP. Thanks to the label provided by the dataset, it is possible to know the beginning and the end of each activity. As previously described, an event $e_i$ is composed of the sensor ID $s_i$, the value $v_i$ and the timestamp $t_i$. By concatenating the sensor ID $s_i$ with his value $v_i$ and by ignoring the timestamp $t_i$, for the reasons explained previously, a sensor word is created, e.g., $s_i = M001$ and the value $v_i = ON$ becomes $M001ON$. All these different text words define the smart home vocabulary to describe activities.

Then, as in NLP, each word in the sequences are transformed into an index to be usable by a neural network. In NLP the index starts at 1, the 0 value is reserved for the sequence padding. Indexes are assigned based on word frequency, e.g., if the word $M001ON$ has the highest occurrence in the dataset, the assigned index is the lowest one i.e $1$.

Sequences are then passed through an embedding layer which transforms index tokens (words) into auto learned features vectors. This creates a simple word embedding that helps the network to get an internal representation of each word in our cases each sensor event.

\subsection{FCN Structure} \label{sec:fcn}
The FCN is a particular CNN. Its structure only contains convolutional layers e.g., no fully connected layers for the classification part. The same structure as \cite{wang2017time,fawaz2019deep} is used in this paper.

\begin{figure} [thpb]
    \centering
    \includegraphics[scale=0.8]{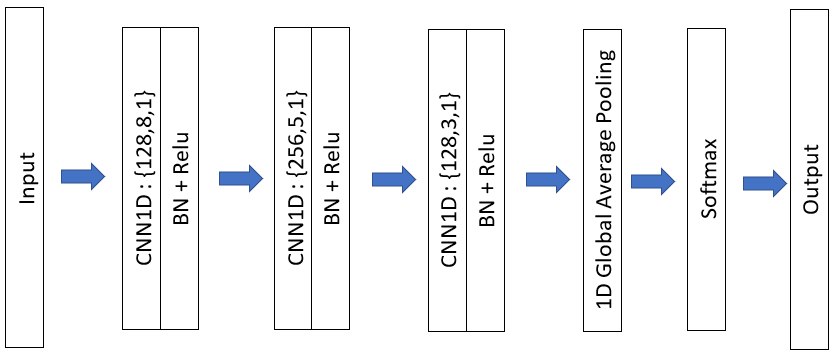}
    \caption{Fully Convolutional Network (FCN) model core}
    \label{fig:FCN_archi_classif}
\end{figure}

This structure (Figure \ref{fig:FCN_archi_classif}) is composed of three blocks described by EQ \ref{eqn:fcn_block}. Where $x$ is the input, $W$ the weight matrix, $b$ the bias and $\otimes$ the convolution operator and $h$ the hidden representation. Each block consists of a 1D convolutional layer with Batch Normalization (BN) \cite{ioffe2015batch} and a rectified linear unit (ReLU) activation to speed up the convergence and help improve generalizations.

\begin{equation}\label{eqn:fcn_block}
\begin{split}
y = W \otimes x + b \\
z = BN(y) \\
h = ReLU(z) \\
\end{split}
\end{equation}
After the three convolution blocks, features are fed into a Global Average Pooling (GAP) layer \cite{lin2013network}. GAP is a pooling operation designed to replace fully connected layers in classical CNNs. The idea is to generate one feature map for each corresponding category of the classification task. The resulting vector is fed directly into the softmax layer to realize the final classification.

One advantage of GAP over the fully connected layers is that it is more native to the convolution structure by enforcing correspondences between feature maps and categories. Thus, the feature maps can be easily interpreted as category confidence maps. Another advantage is that there is no parameter to optimize in the GAP thus over fitting is avoided at this layer. Furthermore, GAP sums out the spatial information; thus it is more robust to spatial translations of the input.

One of the advantages of FCNs is the invariance in the number of parameters across time series of different lengths. This invariance due to using a GAP layer enables the use of a transfer learning approach where one can train a model on a certain source dataset and fine-tune it on the target dataset \cite{fawaz2018transfer}.

\subsection{Sliding Window}\label{sliding_windows}
Contrary to LSTMs, CNNs  must have a fixed input size and activity sequences can have different lengths, between 1 and more than 5000 events. To tackle this issue, a sliding window is applied over sequences. Using a sliding window also allows anticipating an online HAR. To fill windows with fewer events than the window size, a zero padding is used. The zero padding can impact the final result. To avoid too much zero in the sliding windows, a fine-tuned window size must be found.

For experiments, the Sensor Event Windows (SEW) \cite{quigley2018comparative} was used. The SEW approach divides the data into equal sensor event intervals. The size of a SEW is defined by a number of events. Therefore, the duration of the windows may vary. Authors of \cite{quigley2018comparative} compared different windows types and conclude that Time Windows (TW) provides the best accuracy and F-Measure score. They consider SEW as the second-best window method because SEW are able to classify more activities than TW. We assume this is because SEWs keep a fixed context size while it is variable for TWs. A stable context size allows the neural network to keep the same amount of information regardless of the window.

In this work, SEWs were used for two reasons. First, we want to evaluate the method for its ability to learn automatic features from the window context. The intuition being to train a network onto bounded activity sequences to extract features and then use them on streaming sensor data for online recognition. Second, this avoids too many zeros inside windows by controlling the number of events.

\section{Experimental Setup}
LSTMs provide very good results on sequence problems and go beyond traditional advanced HAR methods in Smart Homes \cite{liciotti_lstm}. In order to evaluate the method, LSTMs and FCNs were compared with two dataset ARUBA and MILAN from the widely spread CASAS \cite{cook2012casas} benchmark datasets.

\subsection{Datasets Description}\label{dataset}
Two datasets, ARUBA and MILAN (Table \ref{table:1}) from CASAS were selected for the experiments. The CASAS datasets were introduced by Washington State University. Daily activities data collected, comes from real apartments and houses with real inhabitants, who live in their own houses. The houses are equipped with temperature and binary sensors, as motion or doors sensors. 

A single person carried out activities in both the datasets. The MILAN dataset was selected for the noise on the dataset produced by the pet, which increases the difficulty of classification. They contain several months of labeled activities and are unbalanced, i.e., some activities are less represented than others. In addition these two datasets contain common and different activities with approximately the same number of sensors.

An unbalanced dataset increases the classification challenge. Indeed if some classes are less represented the system gets fewer examples to find the discriminating features. Moreover some events are unlabeled or unidentified and are tagged under the class name "Other". This class appears between 45\% and 50\% into these datasets. In the literature most researchers remove the class "Other" and balance the dataset by reducing the number of examples for each class.

This method creates a drawback, by ignoring unlabeled events it becomes a fixed classification problem. The system cannot make the difference between a known and an unknown class. This does not allow the system to be able to discover new sequences of activities.

Here the original distribution was kept. The objective was to evaluate the robustness of the method and the models.

\begin{table}[hb]
\centering
\caption{Details of datasets.}
\label{table:1}
\begin{tabular}{lrr}  
\hline
  & Aruba & Milan \\
\hline
Habitants               & 1     & 1 + pet   \\
Number of sensors       & 39    & 33        \\
Number of activities    & 12    & 16        \\
Number of days          & 219   & 82        \\
Average sequence length  & 133   & 87.3      \\
\hline
\end{tabular}
\end{table}

\subsection{SEW Parameters}
As previously described, sequences of activities were segmented in SEWs. Different SEWs sizes, 100, 75, 50, 25, with a stride of one was studied. This stride size allows the HAR process each time a new event is triggered. The goal is to find the best SEW size e.g., the minimal SEW size with the maximal information that allows to discriminate activities sequences with a high F1-score and high-balanced accuracy. The smaller the size of SEW, the faster an activity can be recognized in the case of online HAR.

\subsection{Networks Parameters}
FCNs parameters are the same as \cite{fawaz2019deep}. All convolutions have a stride equal to one with a zero padding to preserve the exact length of the time series after the convolution. The first convolution contains 128 filters and a length equal to 8, followed by a second convolution of 256 filters with a length equal to 5, which in turn is fed to a third and final convolutional layer composed of 128 filters, with a length equal to 3.

LSTMs parameters are the same as \cite{liciotti_lstm}. The LSTM cell is composed of 64 neurons and then followed by a softmax layer for the final classification.

As it is usually made in NLP an embedding layer was added between the raw data and the neural network. The number of neurons was fixed to 64 as it was defined in \cite{liciotti_lstm}. 

\subsection{Hardware and Software Setup}
Experiments were made on a server, with an Intel(R) Xeon(R) CPU E5-2640 v3 2.60 GHz, with 32 CPUs, 128 Go of RAM and a NVIDIA Tesla K80 graphic card. Keras and Tensorflow frameworks were used for the algorithm's implementation. The source code can be found at \url{https://github.com/dbouchabou/Fully-Convolutional-Network-Smart-Homes}.

\subsection{Evaluation Method}
To evaluate the proposed method, datasets were split into two parts: 70 \% for the train and 30\% for the test. These two parts contain a shuffled stratified (over class) number of SEW of each activity. e.g., if the dataset contains 100 windows labeled "Sleeping" after shuffling, the 100 windows are split into two parts: 70 windows for the train set, 30 windows for the test set. The random shuffle helps the algorithm to get a better generalization and representation. The stratified forces both subsets to contain representations of each class.

A stratified (over class) threefold cross-validation procedure is performed on the training set. These three trains and three validation subsets are then used to train and validate algorithms.

During the training phase on each train set, early stop and best model selection methods proposed by the Tensorflow framework was used. These methods stop the training before overfitting and saves the best model of each train. The early stop condition is based on the validation loss value. If the current loss doesn't decrease after $n$ epochs since the last, best model selected (here $n = 20$) the training is interrupted.

The three best trained models (one for each training subset) were evaluated on the test set to calculate the average balanced accuracy and the average weighted F1-score, because datasets are unbalanced.

To accelerate the training time by epoch and because the number of SEW is big, a batch size of 1024 was used for all experiments. No differences were noticed between the batch size evaluation during the tests, the results were similar except in training time.

\section{Experimental Results} \label{sec:results}
\subsection{FCNs and LSTMs Performances}
Table \ref{table:2} and Table \ref{table:3} show the performances of two FCNs and two LSTMs on raw sensor data for the two datasets. Vanilla LSTM, FCN and LSTM, FCN with an embedding layer on different windows size were evaluated. The average balanced accuracy and the average weighted F1-score was computed.
FCN appears to obtain the best  weighted F1-score with and without the embedding onto both datasets. The LSTM is close to or equal to the FCN on a large SEW, greater than 50. Compared to the FCN the LSTM looks to need more events to realize the classification.

From the balanced accuracy point of view, FCNs get best values except on the MILAN dataset when the window size is higher than 50. This decrease in performance is due to the zero padding. Indeed the average sequence length on the MILAN dataset is around 88 events. When the window size is close to or over this average, the performances of the FCN decrease. Some small sequences like "Bed to Toilet" or "Eve Meds" are not classified. This results in a drop in the balanced accuracy score.

As an online HAR is expected in our future work, it is interesting to observe the performance of the method on the small SEW size. The goal is to achieve HAR in as little time as possible, with as few events as possible, to get the most responsive system possible. In this case, the FCN obtained the best values with SEWs of sizes 50 and 25. Performances decrease as the SEW size decreases, but the FCN maintained a high score for balanced accuracy and the F1-score. Performance drops less with FCNs than with LSTMs. It seems that FCNs can generate more relevant automatic features than LSTMs on small sequences, therefore with less information.

\begin{table}[htb]
\centering
\caption{Weighted F1 Score and Balanced Accuracy in Aruba's dataset}
\label{table:2}
\begin{tabular}{lrrrr}  
\hline
Model  & 100 & 75  & 50  & 25 \\
\hline
 & \multicolumn{4}{c}{Weighted avg F1 Score (\%)} \\
LSTM     & 96.67           & 94.67           & 90.67           & 85.00          \\
FCN      & 99.00           & 98.00           & 97.67           & 92.33          \\
LSTM + Embedding & \textbf{100.00} & 99.67           & 98.00           & 90.00          \\
FCN + Embedding  & \textbf{100.00} & \textbf{100.00} & \textbf{100.00} & \textbf{99.00} \\
\hline
 & \multicolumn{4}{c}{Balanced Accuracy (\%)} \\
LSTM     & 81.45             & 76.09          & 71.05          & 83.30          \\
FCN      & 88.85             & 87.41          & 87.08          & 80.32          \\
LSTM + Embedding & 94.55             & 93.61          & 90.20          & 74.81          \\
FCN + Embedding  & \textbf{95.37}    & \textbf{95.07} & \textbf{94.89} & \textbf{92.44} \\
\hline
\end{tabular}
\end{table}

\begin{table}[htb]
\centering
\caption{Weighted F1 Score and Balanced Accuracy in Milan’s dataset}
\label{table:3}
\begin{tabular}{lrrrr}  
\hline
Model  & 100 & 75  & 50  & 25 \\
\hline
 & \multicolumn{4}{c}{Weighted avg F1 Score (\%)} \\
LSTM     & 84.00          & 85.67          & 75.33          & 64.00          \\
FCN      & 77.33          & 93.67          & 88.33          & 83.67          \\
LSTM + Embedding & 98.00          & 97.00          & 93.00          & 73.67          \\
FCN + Embedding  & \textbf{99.00} & \textbf{98.00} & \textbf{97.00} & \textbf{94.33} \\
\hline
 & \multicolumn{4}{c}{Balanced Accuracy (\%)} \\
LSTM     & 62.15          & 64.95          & 55.70          & 43.29          \\
FCN      & 42.24          & 76.41          & 71.82          & 71.34          \\
LSTM + Embedding & \textbf{88.52} & \textbf{86.77} & 82.05          & 59.35          \\
FCN + Embedding  & 84.23          & 86.64          & \textbf{87.83} & \textbf{90.86} \\
\hline
\end{tabular}
\end{table}

\subsection{Training Time}
Table \ref{table:4} and Table \ref{table:5} show the average training time and the average amount of training epochs by SEWs size. On both datasets FCNs realized the shortest time on every SEWs size. The embedding layer allows to reduce the number of epochs and the total training time in the majority of cases. The training time is divided by 2 to 6 with the FCN depending on the window size compared to LSTM. This time saving is explained by the ease of parallelization of calculations of convolutional networks.

\begin{table}[htb]
\centering
\caption{Training time performance and number of epochs training in Aruba's dataset}
\label{table:4}
\begin{tabular}{lrrrr}  
\hline
Model  & 100 & 75  & 50  & 25 \\
\hline
 & \multicolumn{4}{c}{Average epoch number} \\
LSTM     & 242 & 278 & 335 & 256 \\
FCN      & 77  & 71  & 111 & 108 \\
LSTM + Embedding & 161 & 191 & 210 & 161 \\
FCN + Embedding  & 67  & 62  & 71  & 98  \\
\hline
 & \multicolumn{4}{c}{Average training time (HH:MM:SS)} \\
LSTM     & 06:28:42 & 06:43:08 & 06:29:58 & 03:00:26 \\
FCN      & 00:58:00 & 00:52:15 & 01:20:35 & 00:51:27 \\
LSTM + Embedding & 04:45:56 & 04:45:38 & 04:14:35 & 02:02:53 \\
FCN + Embedding  & 01:12:37 & 00:59:42 & 00:57:27 & 00:52:15 \\
\hline
\end{tabular}
\end{table}

\begin{table}[htb]
\centering
\caption{Training time performance and number of epochs training in Milan’s dataset}
\label{table:5}
\begin{tabular}{lrrrr}  
\hline
Model  & 100 & 75  & 50  & 25 \\
\hline
 & \multicolumn{4}{c}{Average epoch number} \\
LSTM     & 274 & 385 & 365 & 324 \\
FCN      & 45  & 101 & 87  & 145 \\
LSTM + Embedding & 255 & 290 & 320 & 183 \\
FCN + Embedding  & 65  & 51  & 52  & 55 \\
\hline
 & \multicolumn{4}{c}{Average training time (HH:MM:SS)} \\
LSTM     & 02:03:43 & 02:11:07 & 01:44:06 & 01:00:10 \\
FCN      & 00:09:39 & 00:20:17 & 00:15:08 & 00:16:50 \\
LSTM + Embedding & 01:57:52 & 01:49:35 & 01:36:56 & 00:35:26 \\
FCN + Embedding  & 00:16:24 & 00:11:42 & 00:10:00 & 00:07:67 \\
\hline
\end{tabular}
\end{table}

\subsection{Encoding Impact}
Tables \ref{table:2} to \ref{table:5} show that the embedding layer improves network performances. Indeed, with the embedding layer networks gain significant performance, 10 percentage points on balanced accuracy in average. Sensor events are transformed into vectors of 64 automatically learned features that allow networks to maintain a high score on small SEWs.

During our experiments, we noticed that the frequency encoding strategy improved performance, unlike random or arbitrary index allocation. We think this ordering helps networks generate discriminators on important events or rare events.

\section{Conclusion}
We have proposed a new method that coupled for the first time FCNs and embedding based on frequency encoding for HAR in smart homes. Our assessment on two datasets shows that:
\begin{itemize}
    \item  The embedding based on frequency encoding significantly improves the performance of LSTM and FCN in all cases. This means that the domain knowledge incorporated in the embedding can improve the understanding of events by LSTM and FCN.  
    \item With the same encoding, FCNs obtain the same or better performance than LSTMs, with the exception of only two configurations and are quicker to train. 
    \item Moreover, FCNs outperform LSTMs when the window size decreases. This means that FCNs have a shorter delay in recognizing activities, and are more suitable for real-time activity recognition.
\end{itemize}

The proposed framework is pure end-to-end without any heavy pre-processing on the raw data or feature crafting, thanks to frequency-based encoding and the embedding. This method appears to be relevant for HAR problems in smart homes with low-level sensors.

\section{Discussion and Future Directions}
The results presented in this paper show that the applied DL approach based on NLP encoding and FCN is a relevant solution to significantly improve the smart homes HAR task. 

We used a naive embedding based on frequency encoding that improved classification results. We plan to explore more word embedding techniques \cite{li2020survey} such as Word2Vec or ELMo to improve the latent knowledge space and in the process enhance classification performances. Indeed these techniques take into account the context of words. 
%Most likely, this latent space could contain more knowledge to help classifier algorithms to extract more relevant features on smaller windows.

In addition, we are only experimenting with offline HAR. But the usage of SEWs in our assessment showed relevant results so we want to apply this to online HAR applications. 

Moreover we plan to evaluate other windowing methods as TW or Fuzzy Windows \cite{hamad2019efficient} with this method. To study which window methods produce the fastest and most accurate online HAR in smart homes.
\bibliographystyle{splncs04}
\bibliography{ref.bib}

\begin{thebibliography}{10}
\providecommand{\url}[1]{\texttt{#1}}
\providecommand{\urlprefix}{URL }
\providecommand{\doi}[1]{https://doi.org/#1}

\bibitem{cook2012casas}
Cook, D.J., Crandall, A.S., Thomas, B.L., Krishnan, N.C.: Casas: A smart home
  in a box. Computer  \textbf{46}(7),  62--69 (2012)

\bibitem{fawaz2018transfer}
Fawaz, H.I., Forestier, G., Weber, J., Idoumghar, L., Muller, P.A.: Transfer
  learning for time series classification. In: 2018 IEEE International
  Conference on Big Data (Big Data). pp. 1367--1376. IEEE (2018)

\bibitem{fawaz2019deep}
Fawaz, H.I., Forestier, G., Weber, J., Idoumghar, L., Muller, P.A.: Deep
  learning for time series classification: a review. Data Mining and Knowledge
  Discovery  \textbf{33}(4),  917--963 (2019)

\bibitem{gochoo2018unobtrusive}
Gochoo, M., Tan, T.H., Liu, S.H., Jean, F.R., Alnajjar, F.S., Huang, S.C.:
  Unobtrusive activity recognition of elderly people living alone using
  anonymous binary sensors and dcnn. IEEE journal of biomedical and health
  informatics  \textbf{23}(2),  693--702 (2018)

\bibitem{hamad2019efficient}
Hamad, R.A., Hidalgo, A.S., Bouguelia, M.R., Estevez, M.E., Quero, J.M.:
  Efficient activity recognition in smart homes using delayed fuzzy temporal
  windows on binary sensors. IEEE journal of biomedical and health informatics
  \textbf{24}(2),  387--395 (2019)

\bibitem{hussain2019different}
Hussain, Z., Sheng, M., Zhang, W.E.: Different approaches for human activity
  recognition: A survey. arXiv preprint arXiv:1906.05074  (2019)

\bibitem{ioffe2015batch}
Ioffe, S., Szegedy, C.: Batch normalization: Accelerating deep network training
  by reducing internal covariate shift. arXiv preprint arXiv:1502.03167  (2015)

\bibitem{kowsari2019text}
Kowsari, K., Jafari~Meimandi, K., Heidarysafa, M., Mendu, S., Barnes, L.,
  Brown, D.: Text classification algorithms: A survey. Information
  \textbf{10}(4), ~150 (2019)

\bibitem{li2020survey}
Li, Q., Peng, H., Li, J., Xia, C., Yang, R., Sun, L., Yu, P.S., He, L.: A
  survey on text classification: From shallow to deep learning. arXiv e-prints
  pp. arXiv--2008 (2020)

\bibitem{liciotti_lstm}
Liciotti, D., Bernardini, M., Romeo, L., Frontoni, E.: A sequential deep
  learning application for recognising human activities in smart homes.
  Neurocomputing  (04 2019). \doi{10.1016/j.neucom.2018.10.104}

\bibitem{lin2013network}
Lin, M., Chen, Q., Yan, S.: Network in network. arXiv preprint arXiv:1312.4400
  (2013)

\bibitem{long2015fully}
Long, J., Shelhamer, E., Darrell, T.: Fully convolutional networks for semantic
  segmentation. In: Proceedings of the IEEE conference on computer vision and
  pattern recognition. pp. 3431--3440 (2015)

\bibitem{mohmed2020employing}
Mohmed, G., Lotfi, A., Pourabdollah, A.: Employing a deep convolutional neural
  network for human activity recognition based on binary ambient sensor data.
  In: Proceedings of the 13th ACM International Conference on PErvasive
  Technologies Related to Assistive Environments. pp.~1--7 (2020)

\bibitem{quigley2018comparative}
Quigley, B., Donnelly, M., Moore, G., Galway, L.: A comparative analysis of
  windowing approaches in dense sensing environments. Proceedings
  \textbf{2}(19), ~1245 (Oct 2018). \doi{10.3390/proceedings2191245},
  \url{http://dx.doi.org/10.3390/proceedings2191245}

\bibitem{sedky2018evaluating}
SEDKY, M., HOWARD, C., Alshammari, T., Alshammari, N.: Evaluating machine
  learning techniques for activity classification in smart home environments.
  International Journal of Information Systems and Computer Sciences
  \textbf{12}(2),  48--54 (2018)

\bibitem{singh2017convolutional}
Singh, D., Merdivan, E., Hanke, S., Kropf, J., Geist, M., Holzinger, A.:
  Convolutional and recurrent neural networks for activity recognition in smart
  environment. In: Towards integrative machine learning and knowledge
  extraction, pp. 194--205. Springer (2017)

\bibitem{singh2017human}
Singh, D., Merdivan, E., Psychoula, I., Kropf, J., Hanke, S., Geist, M.,
  Holzinger, A.: Human activity recognition using recurrent neural networks.
  In: International Cross-Domain Conference for Machine Learning and Knowledge
  Extraction. pp. 267--274. Springer (2017)

\bibitem{tahir2020key}
Tahir, S.F., Fahad, L.G., Kifayat, K.: Key feature identification for
  recognition of activities performed by a smart-home resident. Journal of
  Ambient Intelligence and Humanized Computing  \textbf{11}(5),  2105--2115
  (2020)

\bibitem{wang2017time}
Wang, Z., Yan, W., Oates, T.: Time series classification from scratch with deep
  neural networks: A strong baseline. In: 2017 International joint conference
  on neural networks (IJCNN). pp. 1578--1585. IEEE (2017)

\bibitem{yan2019using}
Yan, S., Lin, K.J., Zheng, X., Zhang, W.: Using latent knowledge to improve
  real-time activity recognition for smart iot. IEEE Transactions on Knowledge
  and Data Engineering  (2019)

\end{thebibliography}

\end{document}